\documentclass{revtex4}

\usepackage{graphicx}
\usepackage{amsmath,amssymb}
\usepackage{bm}
\def\beq{\begin{displaymath}}
\def\eeq{\end{displaymath}}
\def\bq{\begin{equation}} 
\def\eq{\end{equation}}
\def\beqn{\begin{eqnarray}}
\def\eeqn{\end{eqnarray}}
\def\m{m_t}

\begin{document}
\bibliographystyle{revtex}

\preprint{MADPH-01-1247}

\title{Higgs plus two jet production at LHC and VLHC}



\author{Carlo Oleari}
\email[]{oleari@pheno.physics.wisc.edu}
\author{Dieter Zeppenfeld}
\email[]{dieter@pheno.physics.wisc.edu}
\affiliation{Department of Physics, University of Wisconsin, Madison, WI 
53706, U.S.A.}


\date{\today}

\begin{abstract}
We consider gluon-fusion and weak-boson fusion production of
Higgs plus two jets in $pp$ collisions at $\sqrt{s}=14$~TeV (LHC) and 
$\sqrt{s}=50, 100, 200$~TeV (VLHC). 
We give cross sections for Higgs masses
between 115~GeV and 200~GeV, and discuss the experimental cuts that should
be applied in order to enhance the weak-boson fusion contribution with
respect to the gluon-fusion background.
\end{abstract}

\maketitle

\section{Introduction}
\label{E4_oleari_0927_sec:intro}
Gluon fusion and weak-boson fusion (WBF) are expected to be the most
copious sources of Higgs bosons in $pp$ collisions at the Large Hadron
Collider (LHC) and a future Very Large Hadron
Collider (VLHC). Beyond representing the most promising discovery
processes~\cite{CMS,ATLAS,wbf}, these two production modes are
also expected to provide a wealth of information on Higgs couplings to gauge
bosons and fermions~\cite{Zeppenfeld:2000td}. The extraction of Higgs boson
couplings, in particular, requires precise predictions of production cross
sections.

A key component of the program to measure Higgs boson couplings at a $pp$
collider is the WBF process, $qq\to qqH$ via $t$-channel $W$ or $Z$ exchange, 
characterized by two forward quark jets~\cite{wbf,Zeppenfeld:2000td}.
QCD radiative corrections to WBF are known to be small~\cite{WBF_NLO} and,
hence, this process promises small systematic errors. 
$H+2$~jet production via gluon fusion, while part of the inclusive Higgs
signal, constitutes a background when trying to isolate the $HWW$ and $HZZ$
couplings responsible for the WBF process.

In two recent papers~\cite{DKOSZ} we presented the details of the
calculation and some results for the real-emission corrections to gluon
fusion which lead to $H+2$~parton final states, at order $\alpha_s^4$, at LHC
energies. The contributing subprocesses include quark-quark scattering which
involves top-quark triangles, quark-gluon scattering processes which are
mediated by top-quark triangles and boxes, and $gg\to Hgg$ which
requires pentagon diagrams in addition.  In these papers we investigated the
validity of the large top-mass limit ($\m\to\infty$) 
by comparing a few distributions computed using
the heavy-top effective Lagrangian~\cite{kauffman} with the
corresponding ones computed keeping $\m$ finite.
In addition, we studied the renormalization and factorization
scale-dependence of the resulting $H+2$~jet cross section, and discussed some
phenomenologically important distributions at the LHC.  

In this contribution, we extend the analysis to the higher energy $pp$ 
collisions of a VLHC, at center of mass energies $\sqrt{s}=50,\;100$ and 
200~TeV. In particular, we 
investigate the cuts that need to be applied in order to enhance WBF 
signals with respect to gluon fusion contributions.

\section{${{\bm H+2}}$ jet production at LHC and VLHC}
\label{E4_oleari_0927_sec:VLHC}
The gluon-fusion processes at ${\cal O}(\alpha_s^4)$, together with
weak-boson fusion,
are expected to be the dominant sources of $H+2$~jet events at
the LHC and VLHC. 
The relative size of the two contributions decides the impact which 
gluon-fusion will have on the study of WBF processes.
The gluon-fusion cross sections diverge as the final-state partons
become collinear with one another or with the incident beam directions, or as
final-state gluons become soft. A minimal set of cuts on the final-state
partons is required to define a finite $H+2$~jet cross section, and these
cuts should at the same time anticipate detector capabilities and jet 
finding algorithms.

We consider two broadly defined sets of cuts: a selection for generic 
$H+2$~jet events (called ``inclusive cuts'') and a more stringent selection 
which is typical for WBF studies (``WBF cuts''). 
In going from LHC to the higher VLHC 
energies, harder jet cuts will have to be imposed, 
in particular on the jet transverse momentum, $p_{Tj}$.  We 
distinguish between LHC and VLHC specific cuts: 
\begin{itemize}
\item { Inclusive cuts}
 \beqn   
    \label{E4_oleari_0927_sec_eq:incl_lhc} 
    {\rm LHC: } && p_{Tj}>20~{\rm GeV}, \qquad |\eta_j|<5, \qquad
        R_{jj}>0.6,  \\
    \label{E4_oleari_0927_sec_eq:incl_vlhc} 
    {\rm VLHC:} && p_{Tj}>30~{\rm GeV}, \qquad |\eta_j|<6, \qquad
        R_{jj}>0.6,  
 \eeqn
\item
{ WBF cuts}: previous inclusive cuts plus~\cite{wbf}
  \beqn
    \label{E4_oleari_0927_sec_eq:WBF_lhc} 
    {\rm LHC: } && |\eta_{j_1}-\eta_{j_2}|>4.2,  \qquad
        \eta_{j_1}\cdot\eta_{j_2}<0, \qquad m_{jj}>600~{\rm GeV},\\
    \label{E4_oleari_0927_sec_eq:WBF_vlhc} 
    {\rm VLHC:} && |\eta_{j_1}-\eta_{j_2}|>5, \phantom{.2} \qquad
        \eta_{j_1}\cdot\eta_{j_2}<0, \qquad  m_{jj}>3~{\rm TeV}.
   \eeqn
\end{itemize}

In general, the gluon fusion differential cross section tends to peak for
central jets and and at a relatively small invariant mass, $m_{jj}$, of 
the dijet system. This is because final-state gluons
tend to be soft and the energy of initial gluons is restricted due to the 
rapid fall-off of the gluon distribution, $g(x,\mu_f)$, 
with increasing $x$. The WBF cuts then enhance the WBF contribution with
respect to gluon fusion. For the inclusive cuts, which allow 
soft events, the gluon fusion cross section dominates over WBF
by a factor 3--10. The WBF cuts force the two tagging jets to be well 
separated in pseudorapidity, $\eta$, they must reside in opposite detector
hemispheres and they must possess a large dijet invariant mass. The 
combination of these requirements forces the gluon fusion cross section below 
the WBF rate for all machine energies.

%
%
\begin{table}
\caption{Total cross section, in picobarn, for $H+2$ jet production, in $pp$
collisions at LHC ($\sqrt{s}=14$~TeV) and VLHC ($\sqrt{s}=50, 100, 200$~TeV)
for three different values of the Higgs mass $m_H$. Results are shown for
gluon fusion (GF) and WBF, using the inclusive cuts of
Eq.~(\ref{E4_oleari_0927_sec_eq:incl_lhc}) and the WBF cuts of
Eqs.~(\ref{E4_oleari_0927_sec_eq:incl_lhc})
and~(\ref{E4_oleari_0927_sec_eq:WBF_lhc}) for LHC, and the inclusive cuts of
Eq.~(\ref{E4_oleari_0927_sec_eq:incl_vlhc}) and the WBF cuts of
Eqs.~(\ref{E4_oleari_0927_sec_eq:incl_vlhc})
and~(\ref{E4_oleari_0927_sec_eq:WBF_vlhc}) for VLHC.}
\label{E4_oleari_0927_tab:sigma}
\vspace{0.2cm}
\begin{tabular}{c||c|c||c|c||c|c||c|c||c|c||c|c||c|c||c|c||}
    & \multicolumn{8}{c||}{INCLUSIVE CUTS } & 
      \multicolumn{8}{c||}{WBF CUTS} \\  
             \cline{2-17}
 {\large $\sigma$ [pb]}   
   & \multicolumn{2}{c||}{{\phantom{$\Big($} 14 TeV\phantom{$\Big)$}}} 
    &  \multicolumn{2}{c||}{{\phantom{$\Big($} 50 TeV\phantom{$\Big)$}}} 
    & \multicolumn{2}{c||}{100 TeV} &
            \multicolumn{2}{c||}{200 TeV} &
 \multicolumn{2}{c||}{14 TeV} &
 \multicolumn{2}{c||}{50 TeV} &
             \multicolumn{2}{c||}{100 TeV} & \multicolumn{2}{c||}{200 TeV} \\
    & \mbox{\ GF\ }&  WBF & \mbox{\ GF\ }&  WBF & \mbox{\ GF\ } & WBF &
      \mbox{\ GF\ } & WBF & \mbox{\ GF\ } & WBF &  \mbox{\ GF\ }&  WBF & 
      \mbox{\ GF\ } & WBF
      &\mbox{\ GF\ } & WBF \\ \hline 
$m_H=115$~GeV & 9.9 & 3.3 & 66.2 & 16.8 &  210 & 38.5 & 635 & 78.6 & 
0.57 & 1.4 & 0.71 & 2.5 & 3.7 & 7.3 & 15.1 & 16.6
\\ \hline
$m_H=160$~GeV & 7.9 & 2.4 & 55.3  & 13.1 & 189  & 30.6 & 559 & 63.3 & 0.50
 & 1.2 & 0.68 & 2.3 & 3.6 & 6.8 & 14.8 & 15.8
\\ \hline
$m_H=200$~GeV & 6.8 & 1.8 & 51.2 & 10.8 &  170 & 25.5 & 524 & 53.3 & 
0.44 & 1.0 & 0.65  & 2.1 & 3.5 & 6.3 & 14.4 & 14.4
\\\hline
\end{tabular}
\end{table}

This is demonstrated in Table~\ref{E4_oleari_0927_tab:sigma}, where 
the expected $H+2$~jet cross sections are shown for Higgs masses of 
115, 160, and
200 GeV, and for the two sets of cuts introduced previously. 
The gluon fusion cross sections depend only weakly on the Higgs mass.
Cross sections correspond to the sum over all Higgs
decay modes: finite Higgs width effects are included. We use CTEQ4L 
parton-distribution functions~\cite{cteq4l}. The
factorization scale was set to $\mu_f=\sqrt{p_{T1} \, p_{T2} }$ and we fix
$\alpha_s = \alpha_s(M_Z) =0.12$. Different choices for the renormalization
and factorization scales have been discussed in Ref.~\cite{DKOSZ}, where a
strong dependence on the renormalization scale was found for the gluon fusion
cross section.

\begin{figure}[htb]
\centerline{
\includegraphics[width=0.45\textwidth]{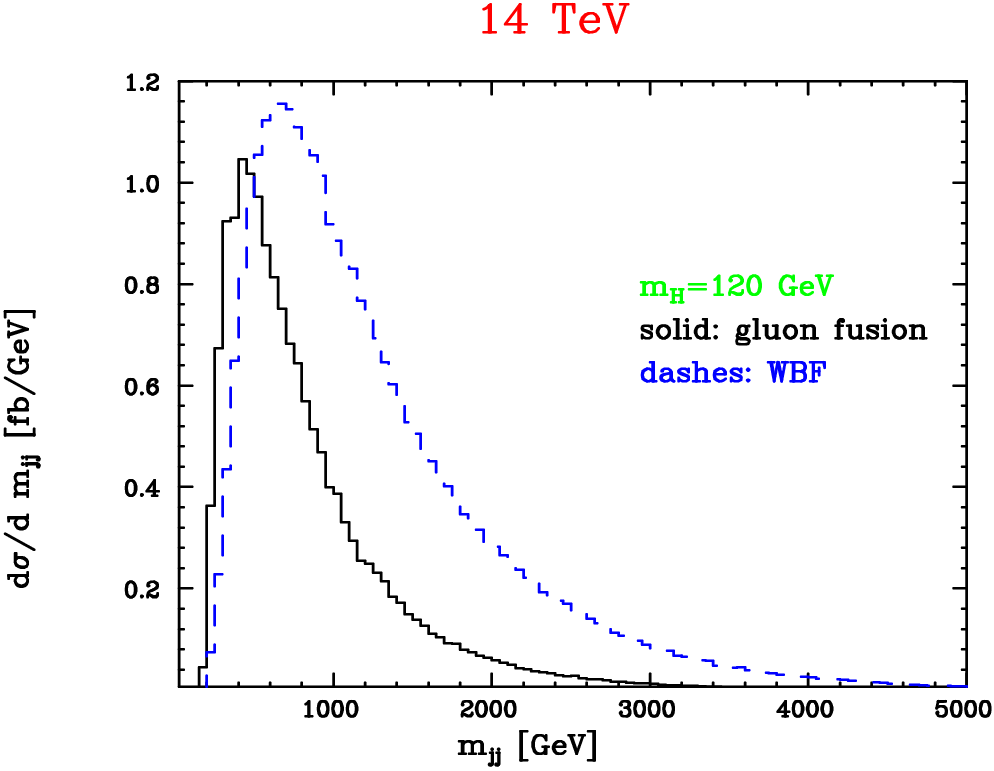}
\includegraphics[width=0.45\textwidth]{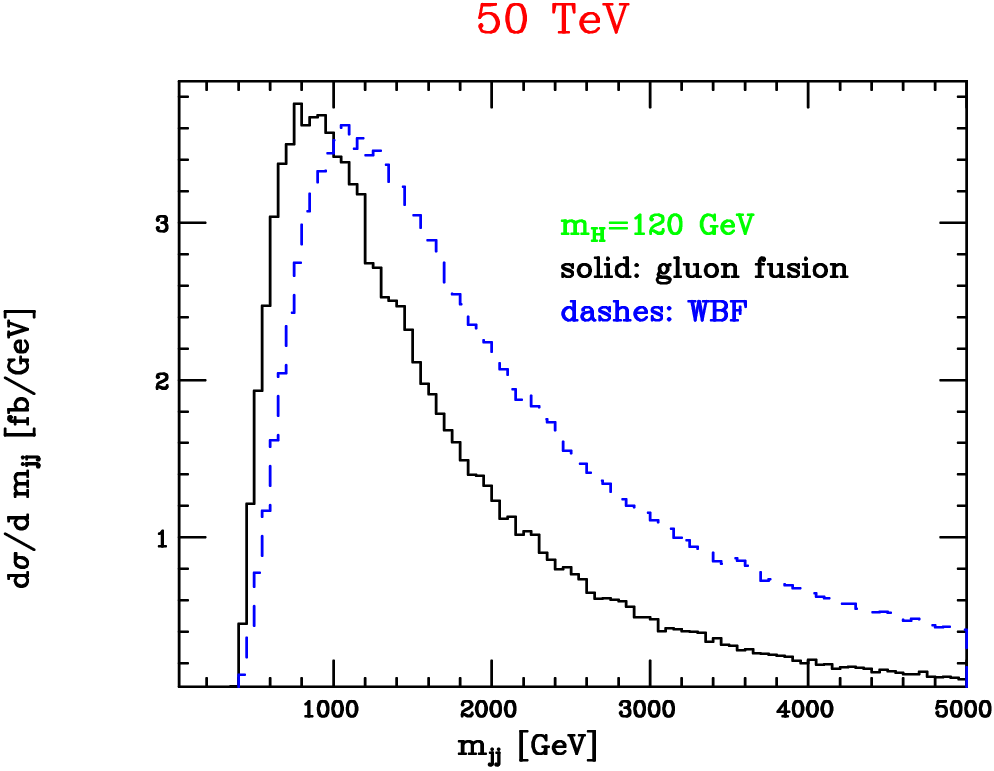}}
\phantom{aaa}
\centerline{
\includegraphics[width=0.45\textwidth]{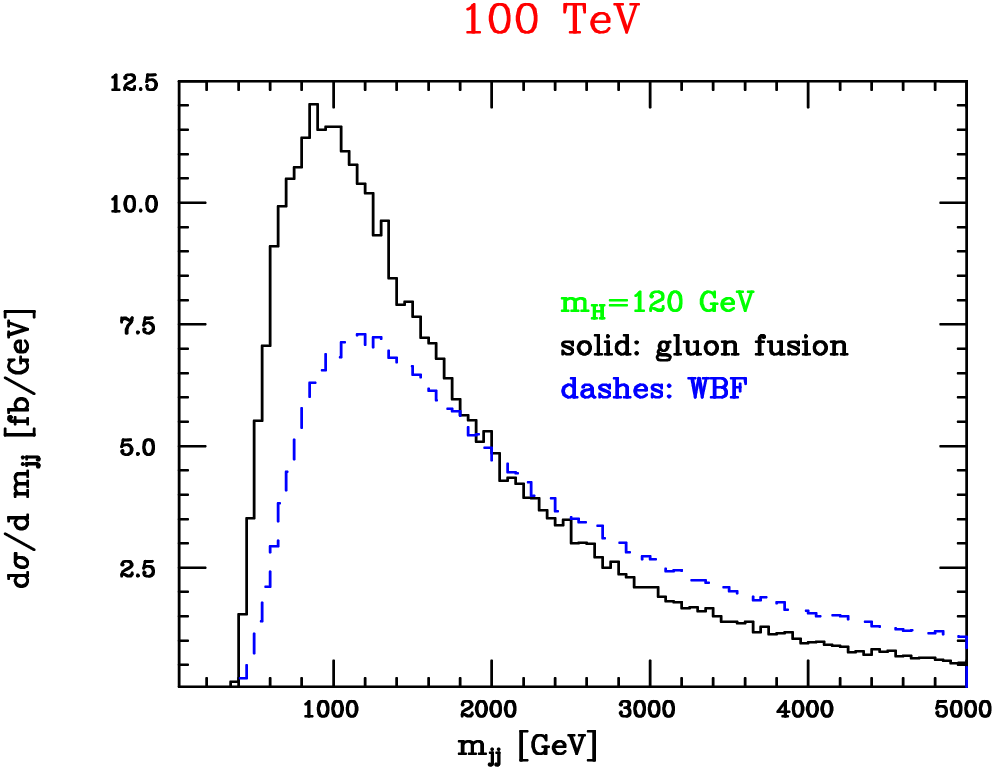}
\includegraphics[width=0.45\textwidth]{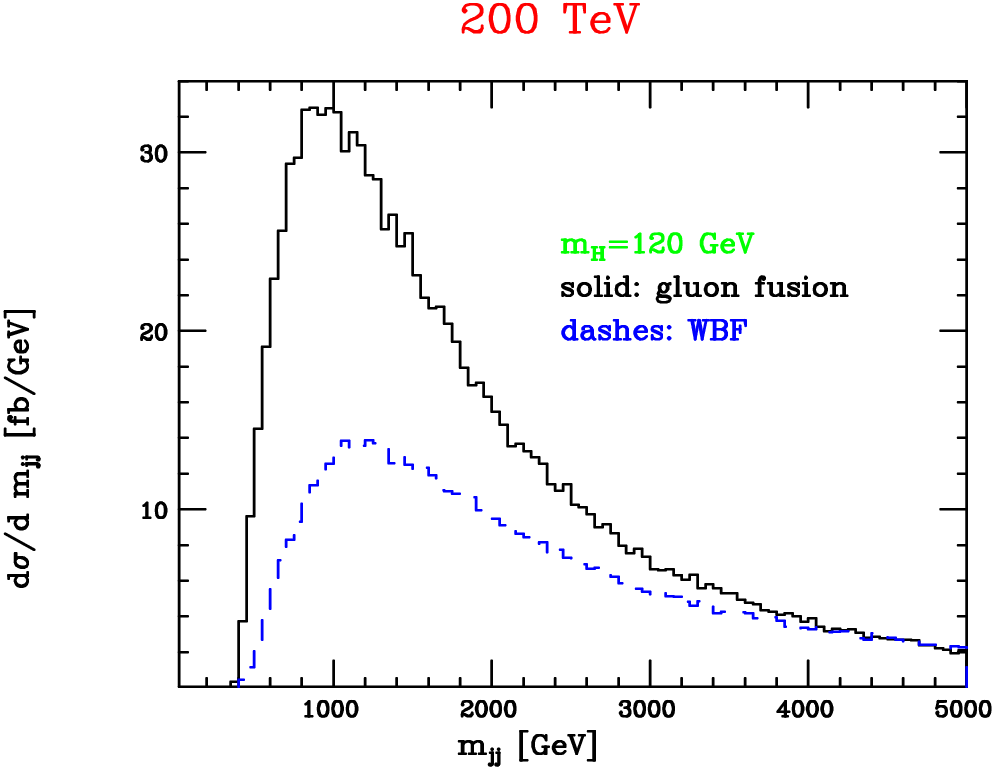}
} 
\caption{Dijet invariant-mass distribution of the two final jets for
gluon-fusion (solid) and WBF (dashes) processes at LHC, $\sqrt{s}=14$~TeV,
and VLHC, with center-of-mass energies of 50, 100 and 200~TeV.  The mass of
the Higgs was fixed at $m_H=120$~GeV. Cross sections are for WBF cuts as 
discussed in the text, without the $m_{jj}$ constraint.}
\label{fig:mjj_WBF_cuts}
\end{figure}

It is instructive to consider the dijet invariant mass distributions for our
two sources of $H+2$~jet events. Within the WBF cuts (but without the 
$m_{jj}$ constraint) they are shown in Fig.~\ref{fig:mjj_WBF_cuts}. At machine 
energies below 100~TeV, WBF soon dominates the $m_{jj}$ distribution: a 
relatively modest $m_{jj}$ requirement (500~GeV at the LHC, 1~TeV at a 
50~TeV machine) will suffice, while retaining most of the WBF signal.
At higher collision energies WBF only dominates at very large dijet invariant
masses, and cuts at 2 or 4~TeV will come at a huge cost of WBF cross section.
In addition, since the jet transverse momenta in WBF events are tied to the
weak boson masses, such cuts leave us with dijet systems with very large 
rapidity separation. Containing such events will require superior rapidity
coverage of VLHC detectors, up to $|\eta|=7$.

\section{Conclusions}

As the center-of-mass energy 
is increased at very high energy hadron colliders, 
the mix of $H+2$~jet events  shifts more and more to 
gluon fusion induced events, 
even with cuts geared towards the enhancement of weak boson fusion.  
A separation of the two sources is partially possible with the rapidity and 
invariant mass cuts discussed above. However, other distinguishing 
characteristics will have to be exploited. Chief among these will be the 
different color structure of the events: $t$-channel color singlet exchange 
for WBF, $t$-channel color octet exchange for gluon fusion. The resulting 
suppression of soft jets in the central region for WBF 
events~\cite{wbf,bjgap} will need to be
studied in detail to prepare tools for the isolation of WBF events at a VLHC.

\begin{acknowledgments}
This summary is based on work done with V.~Del Duca, W.~Kilgore and
C.~Schmidt, whom we thank for enjoyable collaborations and
discussions. C.O.\ acknowledges the support of a DPF Snowmass Fellowship.  This
research was supported in part by the University of Wisconsin Research
Committee with funds granted by the Wisconsin Alumni Research Foundation and
in part by the U.~S.~Department of Energy under Contract
No.~DE-FG02-95ER40896.
\end{acknowledgments}


\end{document}